# Influence of inertial confinement on laser-induced bubble generation and shock wave emission

Xiao-Xuan Liang[1] and Alfred Vogel[1]

1) Institute of Biomedical Optics, University of Luebeck, Peter-Monnik Weg 4, 23562 Luebeck, Germany

Authors to whom correspondence may be addressed:
x.liang@uni-luebeck.de; alfred.vogel@uni-luebeck.de

ORCIDS:  Xiao-Xuan Liang   https://orcid.org/0000-0002-8325-1627

Alfred Vogel   https://orcid.org/0000-0002-4371-9037

## ABSTRACT

Laser-induced breakdown with ultrashort laser pulses is isochoric and inertially confined. It is characterized by a sequence of nonlinear energy deposition and hydrodynamics events such as shock wave emission and cavitation bubble formation. With nanosecond pulses, inertial confinement is lost especially during micro- and nanobubble generation and energy deposition and hydrodynamic events occur concurrently. The onset of bubble expansion during the laser pulse reduces peak pressure, bubble wall velocity, conversion into mechanical energy, and prevents shock wave formation. Here we present an extension of the Gilmore model of bubble dynamics in a compressible liquid that enables to describe the interplay between particle velocity during acoustic transient emission and bubble wall acceleration in the inertial fluid at any degree of confinement. Energy deposition during a finite laser pulse duration is encoded in the time evolution of the bubble's equilibrium radius such that no explicit description of phase transitions is required. The model is used to simulate bubble generation, acoustic transient emission and energy partitioning as a function of laser pulse duration and bubble size at fixed plasma energy density and ambient pressure. It turns out that bubble formation with femtosecond laser pulses is more disruptive than with nanosecond pulses. This applies mainly for micro- and nano-cavitation but to a lesser degree also for millimeter-sized bubbles. We discuss implications for process control in microsurgery and microfluidic manipulation with free-focused laser pulses and via nanoparticle-mediated energy deposition.

**Subject areas:** Laser-Induced Bubbles, Cavitation, Shock Waves, Optical Breakdown, Modeling, Material Processing, Laser Surgery

# I. INTRODUCTION

Laser-induced bubble generation is involved in plasma-mediated laser surgery of cells and transparent tissues [1-7] and is used as a tool for the investigation of fundamental aspects of cavitation bubble dynamics by high-speed photography and acoustic measurements [8-21]. For high-speed photography, mostly millimeter-sized bubbles with relatively long oscillation times are used to provide good spatiotemporal resolution [19,22-25]. By contrast, laser surgery aiming at high precision of the tissue or cellular effects is often associated with the generation of micro- or nanobubbles [3,26-47]. For millimeter-sized bubbles, the oscillation time is four orders of magnitude larger than the laser pulse duration even if nanosecond pulses are used for their generation. However, for microbubbles with much shorter oscillation times the bubble expands significantly during a nanosecond pulse, whereas energy deposition is isochoric only for ultrashort laser pulses. This difference strongly influences pressure evolution and bubble growth during optical breakdown. It needs to be considered in the modeling of bubble expansion and shock wave emission.

Different types of confinement characterize energy deposition in laser-induced bubble formation. "*Thermal confinement*" describes the lack of heat diffusion out of the target volume during the laser pulse. This feature is achieved for all pulse durations $\leq 60$ ns even for tight focusing, where the heat source (plasma) radius $R_0$ may be as small as 0.3 µm [3,48,49]. If not only heat diffusion but also thermal expansion of the heated volume is suppressed during the laser pulse, the energy deposition is isochoric. This condition is often denoted as "*stress confined*" because thermal expansion occurs with sound velocity, and its suppression leads to pressure buildup. Stress confinement is much harder to achieve than thermal confinement. It requires that the laser pulse duration $\tau_L$ is shorter than the stress relaxation time $\tau_{ac}$ given by the propagation time of acoustic waves through plasma volume: $\tau_{ac} = R_0/c_\infty$, where $c_\infty$ is the sound velocity. The buildup of compressive thermoelastic stress results in a radial outward motion of the heated material that, in turn, produces tensile stress within the source, which then facilitates a phase transition and bubble formation [48,50]. Such bipolar stress wave with tensile component is formed if the energy density within the source lies below the kinetic spinodal limit for homogeneous nucleation ($\approx 300°C$ at ambient pressure) [48]. Above this limit, bubble formation is driven by explosive vaporization in the same way as for longer pulses without stress confinement [33]. Thus, thermoelastic tensile stress plays a role only for bubble generation by ultrashort pulse durations in a small energy range reaching from the bubble threshold, $E_{th}$, up to about $1.1 \times E_{th}$ [33]. Since most bubble models including the Gilmore model do not cover bubble nucleation by thermo-elastic tensile stress, we neglect the regime of thermo-elastic bubble formation and focus on the bubble expansion resulting from explosive vaporization, which is always influenced by inertial confinement.

"*Inertial confinement*" refers to the inhibition of radially outward directed mass flow through the inertia of the surrounding medium. We define inertial confinement by the criterion that the volume of the seed bubble, in which the phase transition occurs, increases less than a factor of 2 during the laser pulse. We assume a $\sin^2$ shape of the pulse with duration $\tau_L$ (width at half maximum) and total duration $\tau_L$ [10]. For a spherical source with radius $R_0$, the confinement criterion then reads

$$\left(R\big|_{t=2\tau_L}\right)^3 \leq 2R_0^3, \quad \text{or} \quad \frac{R\big|_{t=2\tau_L}}{R_0} \leq \sqrt[3]{2} \approx 1.26. \tag{1}$$

Fulfillment of this criterion obviously depends on the values of $\tau_L$ and $R_0$, and it depends also on the ambient pressure, $p_\infty$, and the plasma energy density $\varepsilon = E_{abs} / (4/3) \pi R_0^3$, where $E_{abs}$ denotes the pulse energy absorbed within the plasma volume. For given ambient pressure and energy density, both a shorter pulse duration and a larger plasma size increase the confinement degree. An increase of ambient pressure enhances confinement because it inhibits bubble expansion, while an increase of plasma energy density lowers the confinement degree because here the bubble expands faster.

For small bubbles, the resistance against bubble expansion is influenced also by the surface tension $\sigma$ at the bubble wall and the viscosity $\mu$ of the surrounding medium because the pressure arising from these effects scales with $1/R$. More specifically, $P_\sigma = 2\sigma/R$; $P_\mu = 4\mu U/R$, where $U$ is the bubble wall velocity. A schematic of the parameter space influencing laser-induced bubble generation is depicted in Fig. 1.

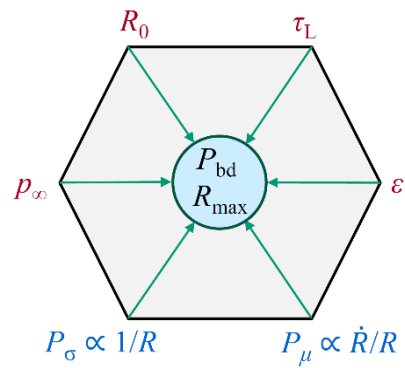

**FIG. 1** Key parameters influencing laser generation of cavitation bubbles. The parameters $\tau_L$, $R_0$, $p_\infty$ and $\varepsilon$ determine the inertial confinement of energy deposition and bubble expansion, while $P_\sigma$ and $P_\mu$, which scale with $1/R$, become additionally relevant for $R_{max} \to 0$.

An investigation of the complex mélange of parameter dependencies influencing laser-induced bubble expansion is of great importance for a deeper understanding of bubble dynamics in general and for parameter optimization in a variety of applications such as laser surgery, microfluidic manipulation, laser ablation in liquids [51], and laser-induced breakdown spectroscopy in water and other liquids [52,53].

For a precise coverage of the initial bubble wall velocity, modeling needs to consider the contributions arising from the bubble pressure and from the particle velocity in the breakdown acoustic transient. The latter can most easily be achieved for isochoric energy deposition by ultrashort laser pulses, where a shock front is immediately formed at the plasma rim and the initial bubble wall velocity is identical with the particle velocity behind the shock front. Liang et al. have modeled the jump-start of bubble expansion for isochoric energy deposition in the framework of the Gilmore model of bubble dynamics in compressible liquids [46]. That approach is here generalized to cover also longer pulse durations without inertial confinement. For considering the time evolution of energy deposition during the laser pulse, we use the approach established by Vogel et al. [10], which has also been employed in Refs. [46,54].

We shall derive the full model for bubble generation with and without inertial confinement in section II and present maps of the inertial confinement degree in dependence on irradiation parameters in section III. The evolution of bubble pressure, bubble wall velocity and maximum bubble radius $R_{max}$ for different degrees of inertial confinement is shown in section IV. In these investigations, ambient pressure and the ratio $R_{nbd}/R_0$ are kept constant, and bubble size and pulse duration are varied to explore the changes for different degrees of inertial confinement. The increasing confinement with decreasing pulse duration results in a rise of breakdown pressure and bubble wall velocity. Section V presents the dependence of the partitioning of absorbed laser energy on $R_{max}$, with special emphasis on the conversion into mechanical energy. Although the influence of viscosity increases for $R_{max} \rightarrow 0$, we will see that an ever larger energy fraction is converted into acoustic energy, when both bubble size and laser pulse duration are reduced. Section VI then shows that for small bubbles a breakdown shock front evolves only with ultrashort laser pulse durations but not during ns breakdown. Altogether, we find that fs breakdown is more disruptive than ns breakdown and that the disruptiveness of laser effects can be adjusted through the choice of the laser pulse duration. In section VII, we discuss the implications of these findings for different applications, with emphasis on laser surgery and laser manipulations in microfluidics.

## II. MODELING

## A. Bubble wall motion interlaced with shock wave emission

In plasma-mediated cavitation, the bubble wall acceleration is driven by the internal bubble pressure but also influenced by the nonlinear sound propagation in the surrounding liquid that involves a pressure-dependent particle velocity. In fs breakdown, energy deposition is isochoric, and the bubble wall does not move during the laser pulse. Here, a shock front forms right at the plasma boundary during the laser pulse. The initial plasma and bubble pressure $P$ is identical with the shock pressure $p_s$, and the initial bubble wall velocity equals the particle velocity behind the shock front. This results in a "jump-start" of the bubble wall. For longer pulse durations, the bubble wall starts to move already during the laser pulse, which influences the pressure buildup during energy deposition. Here, a shock front arises only after the pressure transient has moved a certain distance, which depends on the laser pulse duration as well as on the plasma pressure and size. Nevertheless, the bubble wall exhibits a rapid start due to the particle velocity in the high-pressure liquid region.

Once a shock front is formed, the particle velocity $u_p$ behind the shock is connected to the shock velocity $u_s$ through the jump conditions at the shock front, and the particle velocity can be determined by measuring $u_s$. By doing this for different shock wave pressures in water, Rice and Walsh determined the Hugoniot relation [55]

$$u_p = c_1 \left( 10^{(u_s - c_\infty)/c_2} - 1 \right),$$ (2)

where the constants are $c_1 = 5190$ m/s, $c_2 = 25306$ m/s and $c_\infty$ denotes the sound velocity, $c_\infty = 1483$ m/s. With some rearrangements and the conservation of momentum at a shock front, $p_s - p_\infty = u_s \, u_p \, \rho_\infty$, one obtains a relation between particle velocity and pressure [46,56]:

$$p_s = \rho_\infty u_p \left[ c_\infty + c_2 \log_{10} \left( u_p / c_1 + 1 \right) \right] + p_\infty.$$ (3)

Here $\rho_\infty = 998$ kg/m$^3$ is the mass density of water, and $p_\infty = 10^5$ Pa is the hydrostatic pressure.

We assume that Eq. (3), which links the shock pressure to the particle velocity behind the shock front, approximates the relation between pressure and particle velocity also in a more general fashion and use it to describe the particle velocity at the bubble wall resulting from nonlinear sound propagation at bubble wall pressure $P$. If the second term in the square bracket of Eq. (3) is expressed through its

Taylor expansion and only the first term of the expansion is kept, we obtain [46]

$$P = \rho_\infty c_\infty u_p + \frac{\rho_\infty c_2}{\log(10) c_1} u_p^2 + p_\infty. \tag{4}$$

As shown in [46], resolving this equation for $u_p$ yields

$$u_p = \frac{\sqrt{\rho_\infty^2 c_\infty^2 + 4 \dfrac{\rho_\infty c_2}{\log(10) c_1} P} - \rho_\infty c_\infty}{\dfrac{2 \rho_\infty c_2}{\log(10) c_1}}. \tag{5}$$

We use the Gilmore model of cavitation bubble dynamics [14,46,57] to calculate the temporal development of the bubble radius and the pressure inside the bubble, as well as the pressure distribution in the surrounding liquid. The model considers the compressibility of the liquid surrounding the bubble, viscosity and surface tension. Sound radiation into the liquid from the oscillating bubble is incorporated based on the Kirkwood-Bethe hypothesis and the method of outgoing characteristics [10,46,58-60]. The Gilmore model assumes a constant gas content of the bubble, neglecting evaporation, condensation, gas diffusion through the bubble wall, and heat conduction. Heat and mass transfer influence the bubble pressure at maximum expansion and during collapse [61] but are of little importance for the dynamic behaviour during bubble expansion, which is investigated in the present paper. For large bubble oscillations with strong compression of the contents inside the collapsing bubble, the Gilmore model is often augmented by a van der Waals hard core law to account for a non-compressible volume of the inert gas inside the collapsed bubble [14,46,62]. However, there is no need to consider a van der Waals hard core when studying laser-induced bubble generation.

The bubble dynamics is described by the equation

$$\left(1 - \frac{U}{C}\right) R \dot{U} + \frac{3}{2} \left(1 - \frac{U}{3C}\right) U^2 = \left(1 + \frac{U}{C}\right) H + \frac{U}{C} \left(1 - \frac{U}{C}\right) R \frac{dH}{dR}, \tag{6}$$

where, $R$ is the bubble radius, $U = dR/dt$ is the bubble wall velocity, an overdot means differentiation with respect to time, and $C$ is the speed of sound in the liquid at the bubble wall. $H$ is the enthalpy difference between the liquid at pressure $p(R)$ at the bubble wall and at hydrostatic pressure far away from the bubble $p\big|_{r \to \infty} = p_\infty$

$$H = \int_{p|_{r \to \infty}}^{p|_{r = R}} \frac{dp(\rho)}{\rho}, \tag{7}$$

whereby $\rho$ and $p$ are the density and pressure within the liquid, and $r$ is the distance from the bubble centre. The driving force for the bubble motion is expressed through the difference between the pressure within the liquid at the bubble wall and at a large distance from the wall. Assuming an ideal gas inside the bubble, the pressure $P$ at the bubble wall is given by

$$P = p \left.\right|_{r=R} = \left( p_\infty + \frac{2\sigma}{R_n} \right) \left( \frac{R_n^3}{R^3} \right)^\kappa - \frac{2\sigma}{R} - \frac{4\mu}{R} U \, , \tag{8}$$

where $\sigma$ denotes the surface tension, $\mu$ the dynamic shear viscosity, and $\kappa$ the ratio of the specific heats at constant pressure and volume. The first term on the right hand side describes the gas pressure $P_{gas}$ within the bubble. It is assumed to be uniform throughout the volume of the bubble. The symbol $R_n$ denotes the equilibrium radius of the bubble at which the bubble pressure balances the hydrostatic pressure. The equation of state (EOS) of water is approximated by the Tait equation, with $B = 314$ MPa, and $n = 7$ [63]

$$\frac{p+B}{p_\infty + B} = \left( \frac{\rho}{\rho_\infty} \right)^n . \tag{9}$$

This leads to the following relationships for sound velocity and enthalpy at the bubble wall:

$$C = \sqrt{c_\infty^2 + (n-1)H} \, , \tag{10}$$

$$H = \frac{n(p_\infty + B)}{(n-1)\rho_\infty} \left[ \left( \frac{P+B}{p_\infty + B} \right)^{(n-1)/n} - 1 \right], \tag{11}$$

where $c_\infty$ and $\rho_\infty$ denote the sound velocity and mass density in the liquid at normal conditions. The term $dH/dR$ in Eq. (6) can be derived from Eqs. (8) and (11) by calculating $(dH/dP) \times (dP/dR)$. It reads

$$\frac{dH}{dR} = \frac{1}{\rho_0} \left( \frac{p_\infty + B}{P+B} \right)^{\frac{1}{n}} \times \left( -3\kappa R^2 \left( p_\infty + \frac{2\sigma}{R_n} \right) \frac{\left( R_n^3 \right)^\kappa}{\left( R^3 \right)^{\kappa+1}} + \frac{2\sigma}{R^2} + \frac{4\mu U}{R^2} \right) . \tag{12}$$

The laser pulse is described by a $\sin^2$ function with duration $\tau_L$ (full-width at half-maximum) and total duration $2\tau_L$

$$P_L(t) = P_{L0} \sin^2 \left( \frac{\pi t}{2\tau_L} \right), 0 \le t \le 2\tau_L \, , \tag{13}$$

where $P_L$ denotes the laser power. We model the energy deposition based on the assumption that the cumulated volume increase of the equilibrium bubble at each time $t$ is proportional to the laser pulse energy $E_L$ absorbed up to this time:

$$(4\pi/3)\left[R_n^3(t) - R_0^3\right] \propto E_L(t),$$ (14)

with
$$E_L(t) = \int_0^t P_L(t)dt = P_{L0}\left[\frac{t}{2} - \frac{\tau_L}{2\pi}\sin\left(\frac{\pi t}{\tau_L}\right)\right].$$ (15)

The temporal development of the equilibrium radius $R_n$ during the laser pulse is then [10]

$$R_n(t) = \left\{R_0^3 + \frac{R_{nbd}^3 - R_0^3}{2\tau_L}\left[t - \frac{\tau_L}{\pi}\sin\left(\frac{\pi}{\tau_L}t\right)\right]\right\}^{1/3}$$ (16)

For integrating the rapid start of the bubble wall velocity during the laser pulse into the equation of motion of the Gilmore model, we rewrite Eq. (6) such that it describes the evolution of the bubble wall acceleration and add a term for the time derivative of the particle velocity at the bubble wall

$$\dot{U} = -\frac{3}{2}\frac{U^2}{R}\frac{C - U/3}{C - U} + \frac{H}{R}\frac{C + U}{C - U} + \frac{U}{C}\frac{dH}{dR} + \dot{u}_p$$ (17)

The $\dot{u}_p$ term is obtained through differentiation of Eq. (5):

$$\dot{u}_p = \frac{\dot{P}}{\sqrt{\rho_\infty^2 c_\infty^2 + \dfrac{4\rho_\infty c_2}{\log(10)c_1}P}}.$$ (18)

Since the laser pulse duration is usually significantly shorter than the bubble expansion time, the pressure inside the bubble is at the end of the laser pulse still much larger than the pressure corresponding to surface tension and viscosity. Therefore, we ignore the pressure terms $(-2\sigma/R(t))$ and $(-4\mu U/R(t))$ in Eq. (8) for evaluating $\dot{u}_p$ and focus attention on the compressed gas in the bubble. For an ideal gas and adiabatic conditions with $\kappa = 4/3$, the pressure *evolution* during the laser pulse then becomes

$$P(t) = \left(p_\infty + \frac{2\sigma}{R_n(t)}\right)\frac{R_n^4(t)}{R(t)^4}$$ (19)

*With inertial confinement*, the bubble wall hardly moves while $P(t)$ increases during the laser pulse, and we can set $R(t) = R_0$ for $t \leq 2\tau_L$. Then Eq. (19) changes to

$$P(t) = \left(p_\infty + \frac{2\sigma}{R_n(t)}\right)\frac{R_n^4(t)}{R_0^4},$$ (20)

and the time-derivative of bubble wall pressure reads [46]

$$\dot{P} = \frac{4 p_\infty R_n(t) + 6\sigma}{R_0^4} R_n^2(t) \dot{R}_n(t) .$$ (21)

In the *general case without inertial confinement*, the time-derivative must be obtained directly from Eq. (19) and is more complex

$$\dot{P} = 4\left( p_\infty + \frac{2\sigma}{R_n(t)} \right) \frac{R_n^4(t)}{R(t)^4} \left( \frac{\dot{R}_n(t)}{R_n(t)} - \frac{\dot{R}(t)}{R(t)} \right) - \frac{2\sigma \dot{R}_n(t) R_n^2(t)}{R^4(t)} ,$$ (22)

with $R_n(t)$ from Eq. (16) and

$$\dot{R}_n(t) = \frac{1}{3} R_n^{-2}(t) \frac{R_{nbd}^3 - R_0^3}{2\tau_L} \left[ 1 - \cos\left( \frac{\pi}{\tau_L} t \right) \right] .$$ (23)

A detailed derivation of Eq. (22) is given in the Appendix.

The bubble wall expansion is driven by the gas/vapor pressure inside the bubble as long as this pressure is larger than the sum of hydrostatic pressure and Laplace pressure, $(p_\infty + P_\sigma)$. However, the expansion kinetics is governed not only by the driving pressure but also by the inertia of the fluid mass that gains kinetic energy during expansion. When the driving pressure ceases, the liquid movement becomes purely inertial. For *isochoric energy deposition (inertial confinement)*, the bubble pressure is maximal at the end of the laser pulse and drops continuously afterwards. Here, the transition from pressure-driven to inertia-driven expansion occurs only after the laser pulse and the particle velocity behind the shock front contributes to the bubble wall acceleration during the entire pulse duration. Thus, Eq. (18) becomes

$$\dot{u}_p = \begin{cases} \dfrac{\dot{P}}{\sqrt{\rho_\infty^2 c_\infty^2 + \dfrac{4\rho_\infty c_2}{\log(10) c_1} P}} & \text{for } 0 \le t \le 2\tau_L, \\ 0 & \text{otherwise.} \end{cases}$$ (24)

with $\dot{P}$ from Eq. (21).

By contrast, for long pulses, the pressure rise during energy deposition and the cessation of bubble wall acceleration with transition to inertia-driven dynamics are interlaced. Initially, $\dot{P}$ is positive but $\dot{P}$ becomes negative already during the laser pulse. Because the pressure-related particle velocity in the liquid ceases to accelerate the bubble wall, when $\dot{P} = 0$, it must be considered only for $\dot{P} \ge 0$. Thus, for *energy deposition without inertial confinement* we must modify Eq. (24) by introducing the additional condition $\dot{P} \ge 0$, and we must use $\dot{P}$ from Eq. (22):

$$\dot{u}_{\mathrm{p}} = \begin{cases} \dfrac{\dot{P}}{\sqrt{\rho_\infty^2 c_\infty^2 + \dfrac{4\rho_\infty c_2}{\log(10)c_1}P}} & \text{for } 0 \le t \le 2\tau_{\mathrm{L}} \text{ and } \dot{P} \ge 0, \\[4mm] \qquad\qquad 0 & \text{otherwise.} \end{cases} \tag{25}$$

The above treatment enables to track the interlaced processes of energy deposition, shock wave emission and bubble wall acceleration without explicit modeling of the laser-induced phase transition. This is possible because the energy deposition is encoded in $R_{\mathrm{n}}(t)$, i.e. in the time evolution of a parameter within the framework of the Gilmore model that already assumes a gaseous bubble content.

The acoustic or shock wave emission is modeled and simulated as described in section 3.3 of Ref. [46].

## B. Energy deposition and partitioning

Simulations of acoustic transient emission and calculations of the energy partitioning are performed using the tools derived in Ref. [46] for the entire lifetime of laser-generated cavitation bubbles. Here we present a summary of the main equations relevant for the bubble generation and expansion up to $R_{\mathrm{max}}$. The corresponding pathways of energy partitioning are illustrated in Fig. 2.

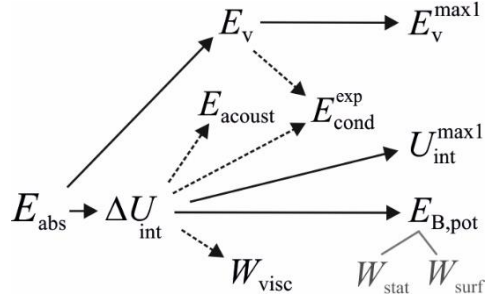

**FIG. 2** Partitioning of the absorbed laser energy $E_{\mathrm{abs}}$ during the first bubble expansion up to $t = t_{\mathrm{max1}}$, when the maximum bubble radius $R_{\mathrm{max}}$ is reached. For explanations see text.

During optical breakdown, the energy absorbed in the plasma volume $V_{\mathrm{P}} = (4/3)\pi R_0^3$ partitions into vaporization energy

$$E_{\mathrm{v}} = \rho_\infty (4/3)\pi R_0^3 \left[ C_{\mathrm{p}}(T_2 - T_1) + L_{\mathrm{V}} \right] \tag{26}$$

and an internal energy gain $\Delta U_{\mathrm{int}}$ of the heated, pressurized gas volume, which drives the bubble expansion. Here $T_1$ and $T_2$ denote the room temperature (20°) and boiling temperature of water (100°C), respectively, $C_{\mathrm{p}}$ = 4187 J/(K·kg) is the isobaric heat capacity of water at 20°C, and $L_{\mathrm{V}}$ = 2256 kJ/kg is the latent heat of vaporization at 100°C. An equation for $\Delta U_{\mathrm{int}}$ will be given further below.

The expanding bubble content does work, $W_{gas}$, on the surrounding liquid, and the internal energy decreases accordingly. The index "gas" refers here to both water vapor and the non-condensable gas produced by plasma-mediated water dissociation. The total energy involved in the bubble expansion is

$$E_{abs} = E_v + \Delta U_{int}(t) + W_{gas}(t).$$ (27)

For isochoric energy deposition with ultrashort laser pulses, $E_{abs} = E_v + \Delta U_{int}$, and the work on the liquid starts only after the end of the laser pulse, when the energy of the free electrons in the laser plasma has been thermalized. In the general case, however, conversion of $\Delta U_{int}$ into $W_{gas}$ starts during the laser pulse and the gas needs to overcome liquid viscosity, the hydrostatic pressure $p_\infty$, and the pressure $p_{surf}$ already during this time. Thereby, it creates kinetic energy $E_{kin}$ of the accelerated liquid, potential energy $E_{B,pot}$ of the expanding bubble, drives the emission of an acoustic transient or shock wave with energy $E_{acoust}$, and does the work $W_{visc}$ by overcoming viscous damping. Thus, the work done by the gas involves the components

$$W_{gas} = E_{acoust} + E_{kin} + W_{visc} + E_{B,pot}, \quad \text{with} \quad E_{B,pot} = W_{stat} + W_{surf},$$ (28)

where $W_{stat}$ and $W_{surf}$ denote the work done against hydrostatic pressure and surface tension.

The change of internal energy up to time $t$ during bubble expansion is given by [46]

$$\Delta U_{int}^{exp}(t) = \frac{4\pi}{3(\kappa-1)}\left\{ \left( p_\infty + \frac{2\sigma}{R_n(t)} \right) \left( \frac{R_n^{3\kappa}(t)}{R^{3\kappa}(t)} \right) R^3(t) - \left( p_\infty + \frac{2\sigma}{R_0} \right) R_0^3 \right\}.$$ (29)

Since the last term on the right hand side of Eq. (29) is very small, we will neglect it in the following. The work done at a given time to overcome the hydrostatic pressure, surface tension, and viscosity is described by [46]

$$W_{stat}(t) = \int p_\infty \, dV = \int 4\pi R^2 \, p_\infty \, dR = \int_0^t 4\pi R(t')^2 \, U(t') \, p_\infty \, dt',$$ (30)

$$W_{surf}(t) = \int p_{surf} \, dV = \int 4\pi R^2 \, p_{surf} \, dR = 8\pi \sigma \int_0^t R(t') U(t') dt',$$ (31)

and

$$W_{visc}(t) = \int p_{visc} \, dV = \int 4\pi R^2 p_{visc} \, dR = 16\pi \mu \int_0^t R(t') U(t')^2 \, dt'.$$ (32)

The transformation of differential variables from $dR$ to $dt'$ in Eqs. (30) to (32) is done using $dR = (dR/dt') \times dt' = U(t')dt'$.

At $R = R_{max}$, the kinetic energy is zero and the potential energy reaches its maximum value. At this stage, the energy of the breakdown acoustic transient, $E_{\text{acoust}}$, can be obtained by evaluating all other terms in Eqs. (27) and (28) and subtracting them from $E_{\text{abs}}$. In this paper, we use the notation "acoustic transient" rather than "breakdown shock wave" because we will see later that no shock front evolves, when inertial confinement is lost. By contrast, for a high degree of confinement and with large laser pulse energy and plasma energy density, a strong shock wave is formed and most of the deposited laser energy is carried away as shock wave energy [64].

The indirect determination of acoustic energy by subtracting all other terms from $E_{\text{abs}}$ is applicable to acoustic transients and shock waves of any strength. This differs from Wang's weakly compressible theory [65] that was adopted in the unified model by Zhang et al. [66]. That approach enables continuous tracking of the energy loss by acoustic radiation during the bubble oscillation but is valid merely for bubble wall movements at low Mach number with a velocity below 250 m/s [65]. Vogel et al. observed that the initial bubble wall velocity after a 10-mJ, 6-ns laser pulse reaches a peak value of 2450 m/s, which decays to 250 m/s within 140 ns [10]. Wang's model is valid only after this initial expansion phase during which most of the shock wave energy is emitted.

We see in Fig. 2 that the bubble energy at $t = t_{\text{max1}}$ comprises not only its potential energy but also remnants of the latent heat used for the vaporization of the liquid in the plasma volume during breakdown and of the internal energy deposited into the vapor:

$$E_{\text{B}} = E_{\text{B,pot}} + U_{\text{int}}^{\text{max1}} + E_{\text{v}}^{\text{max1}}. \tag{33}$$

Although the latter contributions are usually small, they need to be evaluated for establishing a complete balance. The dissipation of internal energy and latent heat during bubble expansion occurs through condensation at the bubble wall and heat conduction into the surrounding liquid. The total condensation loss $E_{\text{cond}}$ is the sum of internal energy loss and release of latent heat

$$E_{\text{cond}}^{\text{exp}} = \Delta U_{\text{int,cond}}^{\text{exp}} + \Delta E_{\text{v}}^{\text{exp}}. \tag{34}$$

In order to quantify the loss of internal energy of the bubble by condensation of vapor at the bubble wall during bubble expansion, we first define the internal energy at $t = t_{\text{max1}}$ for which equilibrium conditions at ambient temperature, i.e. isothermal conditions are assumed:

$$U_{\text{int}}^{\text{max1}}\Big|_{p=p_{\text{v}}} = \frac{4\pi}{3(\kappa-1)} p_{\text{v}} R_{\text{max1}}^3 = 4\pi \, p_{\text{v}} \, R_{\text{max1}}^3. \tag{35}$$

The internal energy loss during bubble expansion is given by the difference between the energy of an adiabatically expanding bubble at $R_{max}$, which is obtained by evaluating Eq. (29) at the time of maximum bubble expansion for $R_n = R_{nbd}$, and the energy corresponding to isothermal conditions at $R_{max}$ from Eq. (35):

$$\Delta U_{int,cond}^{exp} = U_{int}^{exp}\Big|_{R_n = R_{nbd}} - U_{int}^{max1}\Big|_{p = p_v} \tag{36}$$

The internal energy lost by condensation is dissipated as heat into the surrounding liquid. In addition, we need to look at the latent heat released into the liquid. For this purpose, we express the amount of vapor contained in the bubble at different instants in time (directly after the laser pulse, at $R_{max1}$, $R_{min1}$, and $R_{max2}$) through the radius of a vapor bubble at room temperature with pressure 0.1 MPa. The vapor bubble radius after breakdown is calculated considering conservation of mass during vaporization of the liquid in the plasma volume. With

$$\rho_\infty \frac{4}{3}\pi R_0^3 = \rho_v \frac{4}{3}\pi (R_v^{bd})^3, \quad \text{we obtain} \qquad R_v^{bd} = R_0(\rho_\infty / \rho_v)^{1/3}. \tag{37}$$

The mass density of vapor at $p_v = 0.1$ MPa and $T = 20°C$ is $\rho_v = 0.761$ kg/m³. The amount of vapor in the expanded bubble can be assessed by assuming that the vapor pressure at $R_{max1}$ and $R_{max2}$ equals the equilibrium vapor pressure at room temperature, $p_v = 2.33$ kPa [14]. The corresponding bubble radii for vapor at ambient pressure are then

$$R_v^{max\,i} = R_{max\,i}(p_v / p_\infty)^{1/3}, \text{ with i = 1 and 2.} \tag{38}$$

The loss of latent heat by condensation during bubble expansion is given by

$$\Delta E_v^{exp} = E_v - E_v^{max1}, \text{ with} \qquad E_v^{max1} = \left(\frac{R_v^{max1}}{R_v^{bd}}\right)^3 E_v \tag{39}$$

and $E_V$ from Eq. (26).

Evaluation of Eqs. (29) – (32) enables tracking of the energy partitioning from $E_{abs}$ into $\Delta U_{int}$, $W_{gas}$, $W_{visc}$ and $E_{B,pot}$ during the laser pulse and bubble expansion, while evaluation of equations (29) to (39) at $t = t_{max1}$ together with Eqs. (27) and (28) yields the complete energy balance for the expanded bubble. Thus, even though the Gilmore model does not continuously track heat and mass transfer at the bubble wall, we can get information about integral value of $E_{cond}^{exp}$, the energy loss by condensation during bubble expansion. When experimental data on the bubble oscillation times $T_{osci}$ are available, fitting of the model predictions to $T_{osci}$ by variation of $R_n$ can be used to determine the equilibrium bubble radius after breakdown and at the individual collapse states. This way, $E_{\text{cond}}^{\exp}$ can be evaluated not only for the initial bubble expansion but also for the collapse phase from $R_{\text{max}1}$ to $R_{\text{min}1}$ and for the expansion and collapse phases of subsequent oscillations [46].

## C. Simulations

All numerical simulations in this paper are performed using the generalized model of laser-induced bubble generation described above and in Ref. [46]. We study the influence of inertial confinement on laser-induced bubble formation numerically for a large range of laser pulse durations and bubble sizes. The ambient pressure $p_\infty$ is kept constant at 0.1 MPa and the energy density $\varepsilon$ within the seed bubble representing the plasma volume is also mostly kept constant at an intermediate value. For isochoric energy deposition, $\varepsilon$ is reflected by the ratio $R_{\text{nbd}}/R_0$. For convenience, we use this ratio as a general measure of 'deposited energy density,' i.e. also for cases without inertial confinement, where the bubble expands already during the laser pulse, which lowers the actual energy density. Since we keep $R_{\text{nbd}}/R_0$ constant, $R_0$ has to be used as free parameter for achieving specific $R_{\text{max}}$ values.

Most simulations are performed for $R_{\text{nbd}}/R_0 = 10.4$ taken from Ref. [46]. This value corresponds to a moderate plasma temperature of $\approx 1550$ K, which is well above the bubble threshold but still in the lower part of the large range of energy densities that have been experimentally observed for optical breakdown with tightly focused laser pulses [64]. The choice of a constant $R_{\text{nbd}}/R_0$ value for all simulations provides a convenient reference point for the comparison of bubble dynamics and shock wave emission with different degrees of inertial confinement. Moreover, it is justified by the observation that well above the bubble threshold $\varepsilon$ is approximately constant over a large range of pulse energies [46,64]. Close to threshold, the absorption depth is larger than the plasma length, and only a small fraction of the incoming light is absorbed. Here $\varepsilon$ varies with $E_{\text{L}}$. However, well above $E_{\text{th}}$ the absorption depth of incoming laser light is smaller than the plasma length and the breakdown front moves upstream towards the incoming laser beam during the laser pulse [56,67-70]. Here an increase of $E_{\text{L}}$ results in an enlargement of the plasma volume while the energy density remains approximately constant.

In order to explore the full range of inertial and stress confinement, simulations were performed for laser pulse durations down to 100 fs. This needs to be justified because hydrodynamic phenomena set in only after the energy of the free electrons in the laser-produced plasma is thermalized and a thermodynamic temperature with local thermal equilibrium between electrons and heavy particles has been achieved [71,72]. The thermalization time for plasma in water features a strong dependence on electron density. Well above the bubble threshold, where the band structure of liquid water is altered or even dissolved, thermalization times around 20 ps have been inferred from experiments [73,74]. By contrast, for low-density plasmas below the bubble threshold, the kinetic electron energy can be rapidly dissipated by inelastic collisions involving vibrational excitation of water molecules, and low-energy electrons can then easily hydrate into the network of dipolar water molecules. Here, thermalization times below 300 fs have been derived from experimental investigations [75,76] and through Monte Carlo simulations [77]. Thus, close to the bubble threshold, where the plasma electron and energy density are small, hydrodynamic phenomena such as acoustic transient emission and bubble formation will set in on a sub-picosecond time scale. Since the exact thermalization times in the nano- and microbubble regime close to threshold is not yet known, we cover this regime by performing calculations for pulse durations of 100 fs and 10 ps.

The numerical methods described in this paper and in Ref. [46] are implemented in the open-source software library LIBDAR (**L**aser-**I**nduced **B**ubble **D**ynamics and **A**coustic **R**adiation) that was used to produce the presented results. It is available at https://github.com/X-X-Liang/LIBDAR.

+

## III. DEPENDENCE OF INERTIAL CONFINEMENT DEGREE
## ON IRRADIATION PARAMETERS

Figure 3(a) presents a map of the inertial confinement degree as a function of laser pulse duration and maximum radius of the laser-induced bubble, and Fig. 3(b) illustrates the dependence of initial bubble growth on $\tau_L$ for selected pulse durations. The $R(t)$ curves are shown for bubbles with about 5 µm maximum radius at constant $R_{nbd}/R_0$ ratio.

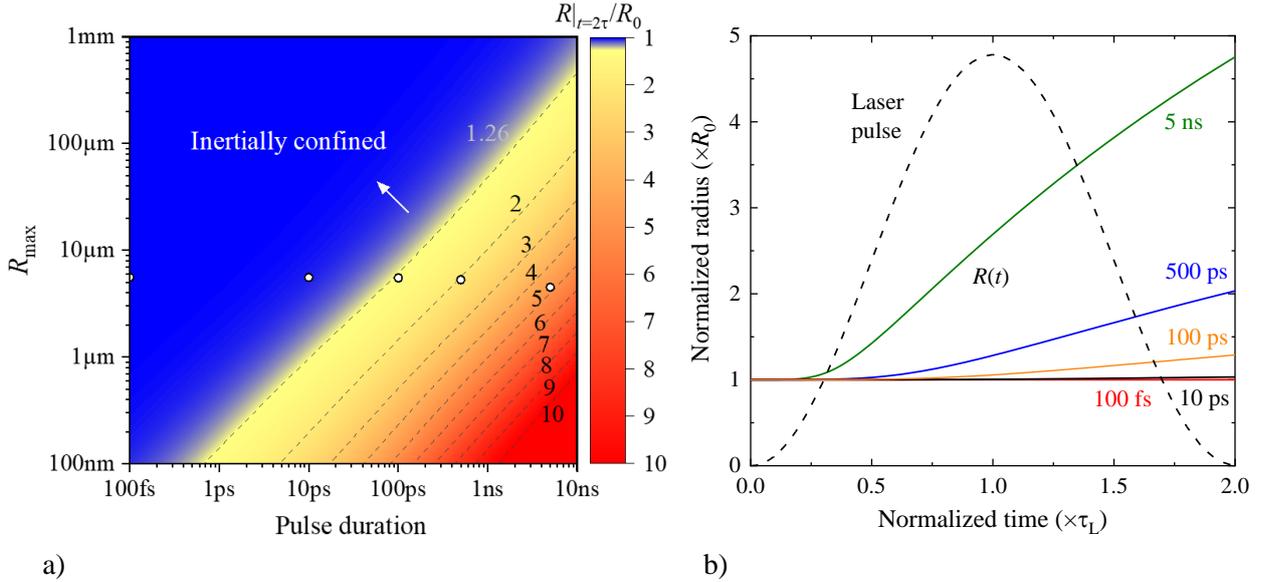

a)                                            b)

**FIG. 3** a) Map of the inertial confinement degree in ($\tau_L$, $R_{max}$) parameter space. b) Bubble expansion during the laser pulse for specific pulse durations. All calculations were performed for $R_0$ = 210 nm and $R_{nbd}$ = 10.4 $R_0$. The ($\tau_L$, $R_{max}$) values for which $R(t)$ curves are displayed in (b) are indicated as dots in (a).

For inertial confinement, the bubble volume at the end of the laser pulse ($t = 2\tau_L$) must be smaller than twice the volume of the laser plasma, which corresponds to $R|_{t=2\tau_L} / R_0 \leq \sqrt[3]{2} \approx 1.26$ [Eq. (1)]. This value and larger ratios up to $R|_{t=2\tau_L} / R_0 = 10$ are indicated as dashed lines in the color-coded confinement map of Fig. 3(a). The decrease of confinement with increasing pulse duration goes along with a dramatic increase of the bubble expansion during the laser pulse, as visible in Fig. 3(b). During a 5-ns pulse, the bubble has already expanded to 22.4% of its maximum radius of 4.4 µm.

Figure 4 shows how the location of the border between inertially confined and non-confined bubble dynamics changes with plasma energy density, which is encoded in the $R_{nbd}/R_0$ ratio. With increasing energy density, the confinement border shifts towards larger bubble sizes and shorter pulse durations [Fig. 4(a)]. As a consequence, micro-and nanobubble generation may be inertially unconfined even with ultrashort laser pulses. Most high-speed photographic studies are performed with pulse durations between 1 ns and 10 ns in the ($R_{max}$, $\tau_L$) region between 100 µm and 5 mm. This region lies in the transition zone between inertially confined and non-confined energy deposition. Therefore, the advanced model presented in this study is not only relevant for nano- and microbubbles but for studies of bubble dynamics in general.

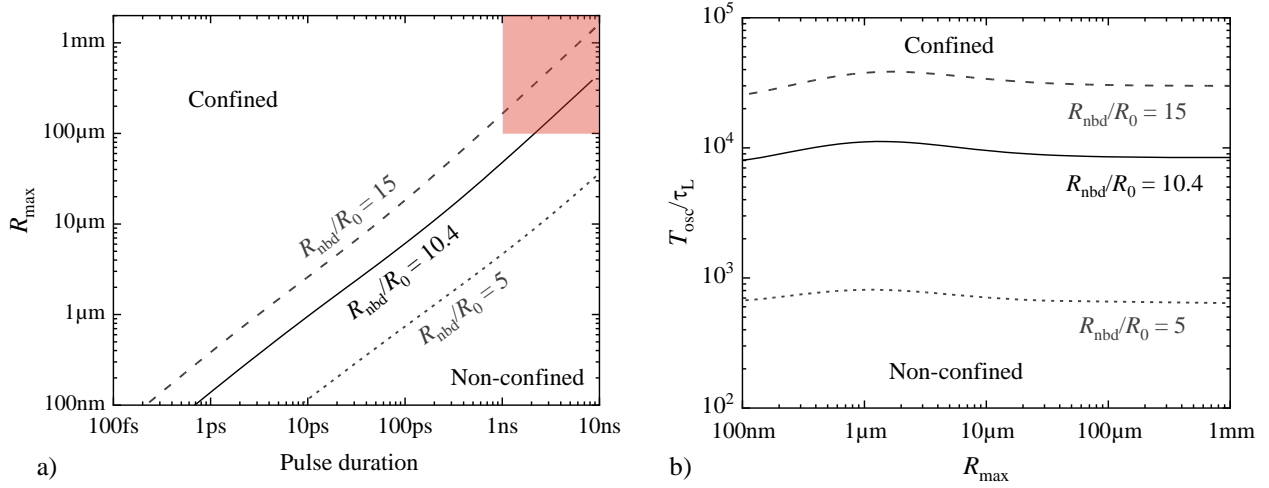

**FIG. 4** (a) Borders of inertial confinement in the ($\tau_L$, $R_{max}$) parameter space for various plasma energy densities, which are here represented by the $R_{nbd}/R_0$ ratio. (b) Ratio between bubble oscillation time and laser pulse duration, $T_{osc}/\tau_L$, at the inertial confinement border for different plasma energy densities.

We see in Fig. 4(b) that with increasing plasma energy density inertial confinement is lost at an ever-larger ratio $T_{osc}/\tau_L$ between bubble oscillation time and laser pulse duration. Already for $R_{nbd}/R_0 = 10.4$ corresponding to a plasma temperature of 1550 K [46], bubble generation is confined only if the oscillation time is more than ≈ 10000 times longer than the laser pulse duration, which for $\tau_L = 5$ ns corresponds to $T_{osc} \geq 50$ µs. For $R_{nbd}/R_0 = 15$, inertial confinement is reached only for $T_{osc}/\tau_L \approx 30000$, i.e. for bubbles with $T_{osc} \geq 150$ µs, corresponding to a maximum radius $R_{max} \geq 0.8$ mm.

Interestingly, the $T_{osc}/\tau_L$ ratio needed for inertial confinement is largely independent of $R_{max}$. For $R_{max} > 10$ µm, this reflects the self-similarity of bubble dynamics in regions, where surface tension and viscosity are negligible. For $R_{max} < 10$ µm, where $P_\sigma$ and $P_\mu$ come into play, the confinement borderline undulates slightly.

## IV. BUBBLE PRESSURE, EXPANSION VELOCITY AND MAXIMUM SIZE
## IN DEPENDENCE ON PULSE DURATION AND BUBBLE SIZE

Figure 5 presents $P(t)$ and $U(t)$ curves of the initial expansion phase of laser-induced bubbles with about 5 μm maximum radius for different pulse durations and constant $R_{nbd}/R_0$ ratio. The pulse duration dependencies for $R_{max}$, peak bubble pressure $P_{bd}$ and peak bubble wall velocity $U_{bd}$ are shown in Fig. 6.

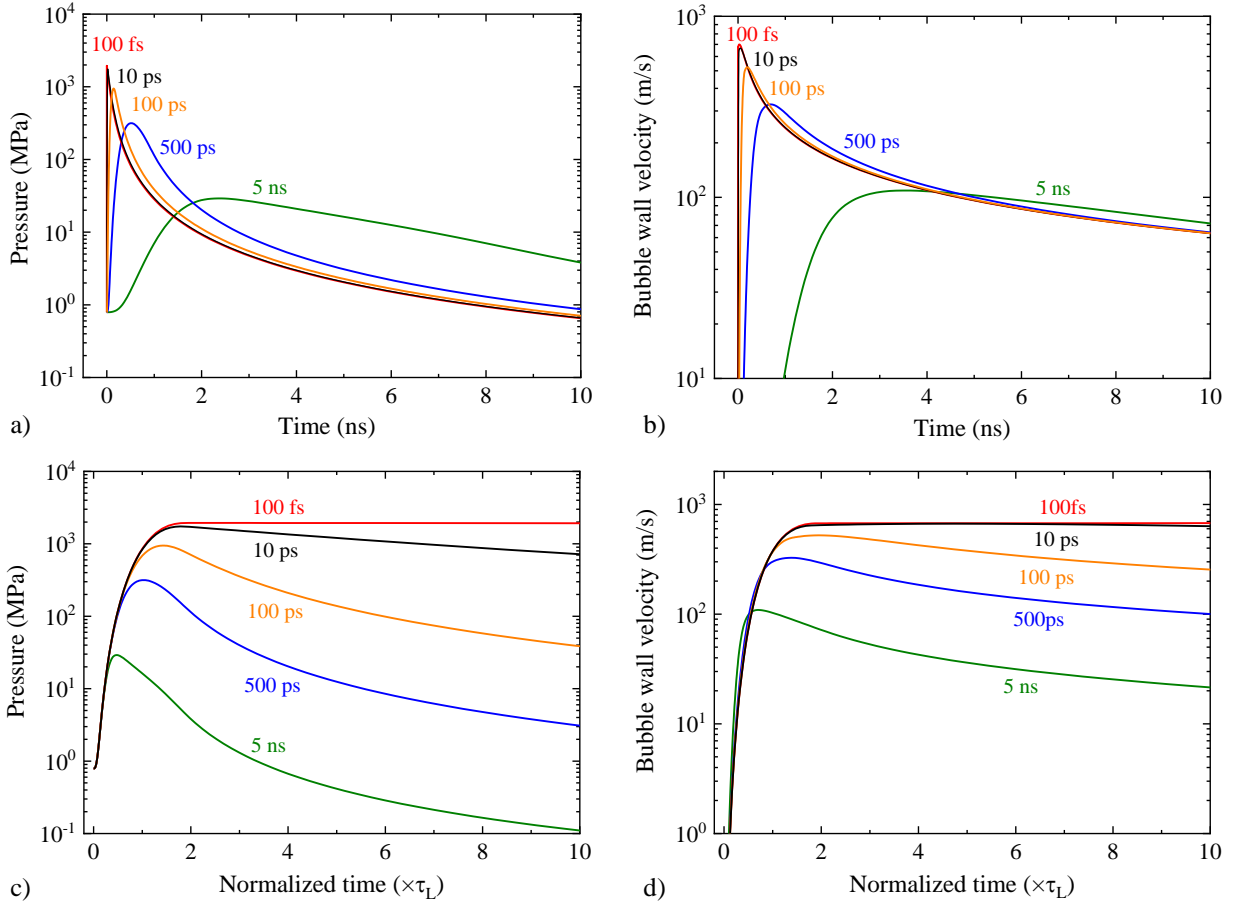

**FIG. 5** Time evolution of the bubble pressure $P_{gas}(t)$ and bubble wall velocity $U(t)$ of bubbles produced by laser pulses with different durations. The time axis is linear in (a) and (b) but normalized with the laser pulse duration in (c) and (d). All simulations were performed for $R_0 = 210$ nm and $R_{nbd} = 10.4\ R_0$, which leads to a maximum bubble size of $R_{max} \approx 5$ μm.

We see in Fig. 6 that $R_{max}$ decreases slightly with increasing $\tau_L$ and decreasing confinement, whereas the peak pressure drops by two orders of magnitude from $P_{bd} = 1943$ MPa at 100 fs to $P_{bd} = 29$ MPa at 5 ns. The peak bubble wall velocity drops from $U_{bd} = 673$ m/s at 100 fs to $U_{bd} = 109$ m/s at 5 ns. The view on the initial bubble dynamics in Fig. 5 reveals that after a 100 fs pulse with full inertial confinement pressure and velocity remain close to their peak value during a time interval considerably longer than the laser pulse. By contrast, when inertial confinement is lost, they start to drop already during the pulse. For $\tau_L$ = 5 ns, peak pressure and velocity are already reached after about one quarter of the laser pulse and drop soon after the bubble starts to expand (as can be seen by comparing Figs. 3b and 5c,d).

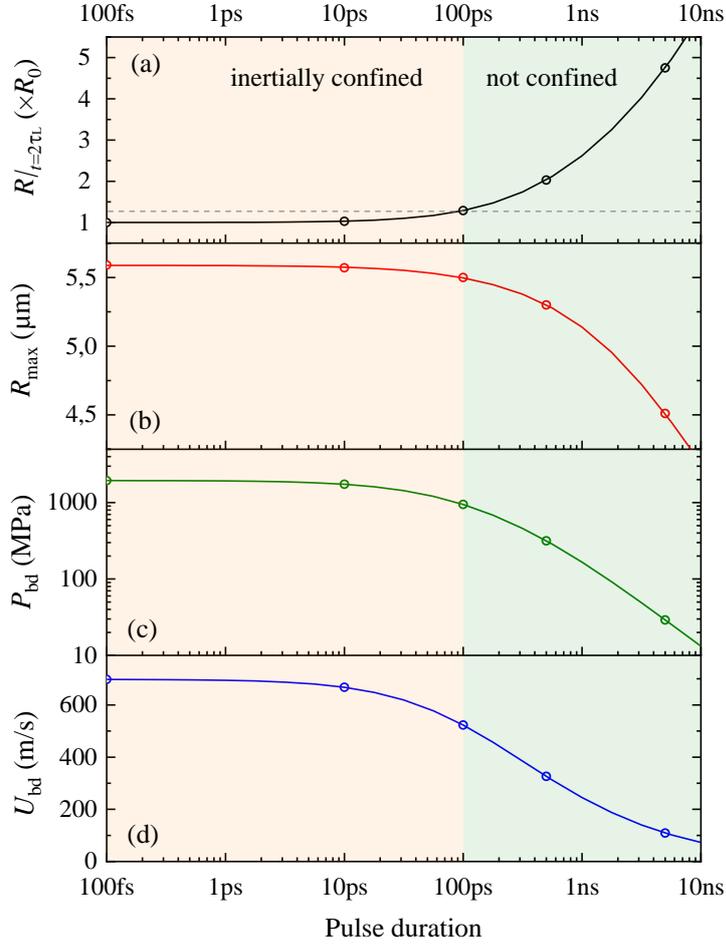

**FIG. 6** Change of bubble parameters as a function of laser pulse duration for $R_0$ = 210 nm and $R_{nbd}$ = 10.4 $R_0$. The loss of inertial confinement degree shown in (a) diminishes the maximum bubble size $R_{max}$ (b), the peak breakdown pressure $P_{bd}$ (c), and the peak bubble wall velocity $U_{bd}$ (d). Calculations were performed for the same pulse durations as in Fig. 5; the solid lines are drawn to guide the eye.

In the next step, we look at the relationship between inertial confinement and bubble size. The bubble size dependence of breakdown pressure and bubble wall velocity is shown in Fig. 7 for different laser pulse durations. For large bubbles, the shock pressure and bubble wall velocity are determined merely by the plasma energy density, which is encoded by $R_{nbd}/R_0$. With decreasing $R_{max}$, the pressure terms $P_\sigma = 2\sigma/R$ from surface tension and $P_\mu = 4\mu U//R$ from viscosity become ever more important. This effect results in an increase of $P_{bd}$ and $U_{bd}$ for $R_{max} \to 0$ and isochoric energy deposition ($\tau_L$ = 100 fs). For longer pulse durations, the increasing influence of surface tension and viscosity is covered by the loss of inertial confinement with decreasing bubble size. Thus, the overall effect at $\tau_L$ = 500 ps and $\tau_L$ = 5 ns is a lowering of $P_{bd}$ and $U_{bd}$ for $R_{max} \to 0$.

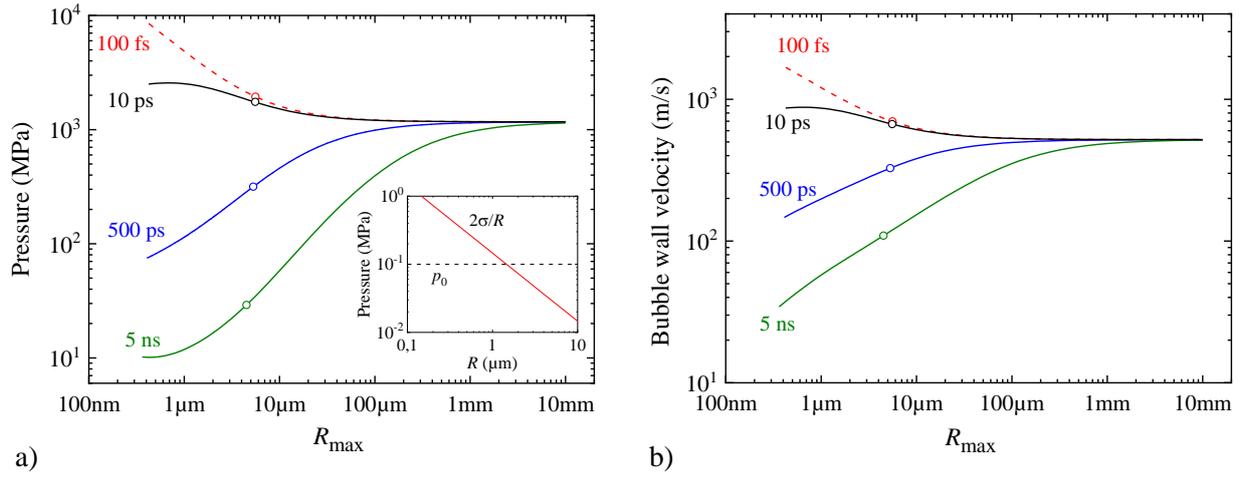

a)                                                    b)

**FIG. 7** Breakdown peak pressure $P_{bd}$ (a) and velocity $U_{bd}$ (b) as a function of $R_{max}$ at constant $R_{nbd}/R_0 =$ 10.4 and different $\tau_L$. The $R_{max}$ values for which data are presented in Fig. 6 are marked as dots. The insert in (a) shows the Laplace pressure from surface tension as a function of bubble radius. Curves for $\tau_L =$ 100 fs are dashed because of the uncertainty of the thermalization time at small plasma energy density.

# V. ENERGY DEPOSITION AND PARTITIONING
# IN DEPENDENCE ON PULSE DURATION AND BUBBLE SIZE

In the present paper, the ratio $R_{nbd}/R_0$ is used as control parameter for the energy density. However, for small bubbles the energy density needed for reaching a given $R_{max}$ value does not only depend on $R_{nbd}/R_0$ but also on $R_{max}$ itself. Overcoming the additional resistance against bubble expansion arising from surface tension and viscosity requires additional energy. For $R_{max} \to 0$, this goes along with a relative increase of seed bubble radius and absorbed energy as shown in Figs. 8(a) and 8(b). Combined, both parameters determine the energy density $\varepsilon = E_{abs}/(4/3)\pi R_0^3$, which is presented in Fig. 8(c). For $\tau_L = 100$ fs, the energy density strongly increases with decreasing bubble size, while it remains approxi mately constant for $\tau_L = 5$ ns. The difference is due to the counteracting trends of the $\tau_L$ dependencies of $R_0(R_{max})$ and $E_{abs}(R_{max})$ in Figs. 8(a) and (b). In this context, it is important to remember that the calculated $\varepsilon$ value represents a real-world entity only for isochoric energy deposition at $\tau_L = 100$ fs. For longer pulse durations, where the bubble already expands during the laser pulse, $\varepsilon$ is a virtual value and the real peak values of energy density within the expanding bubble drops strongly for $R_{max} \rightarrow 0$, like $P_{bd}$ and $U_{bd}$ in Fig. 7. We see in Fig. 8(d) that the ratio $R_{max}/R_0$ drops for $R_{max} \rightarrow 0$, i.e. the maximum bubble size achieved for a given seed bubble radius decreases. This drop is due to losses by viscous damping and to the increase amount of work needed to overcome surface tension.

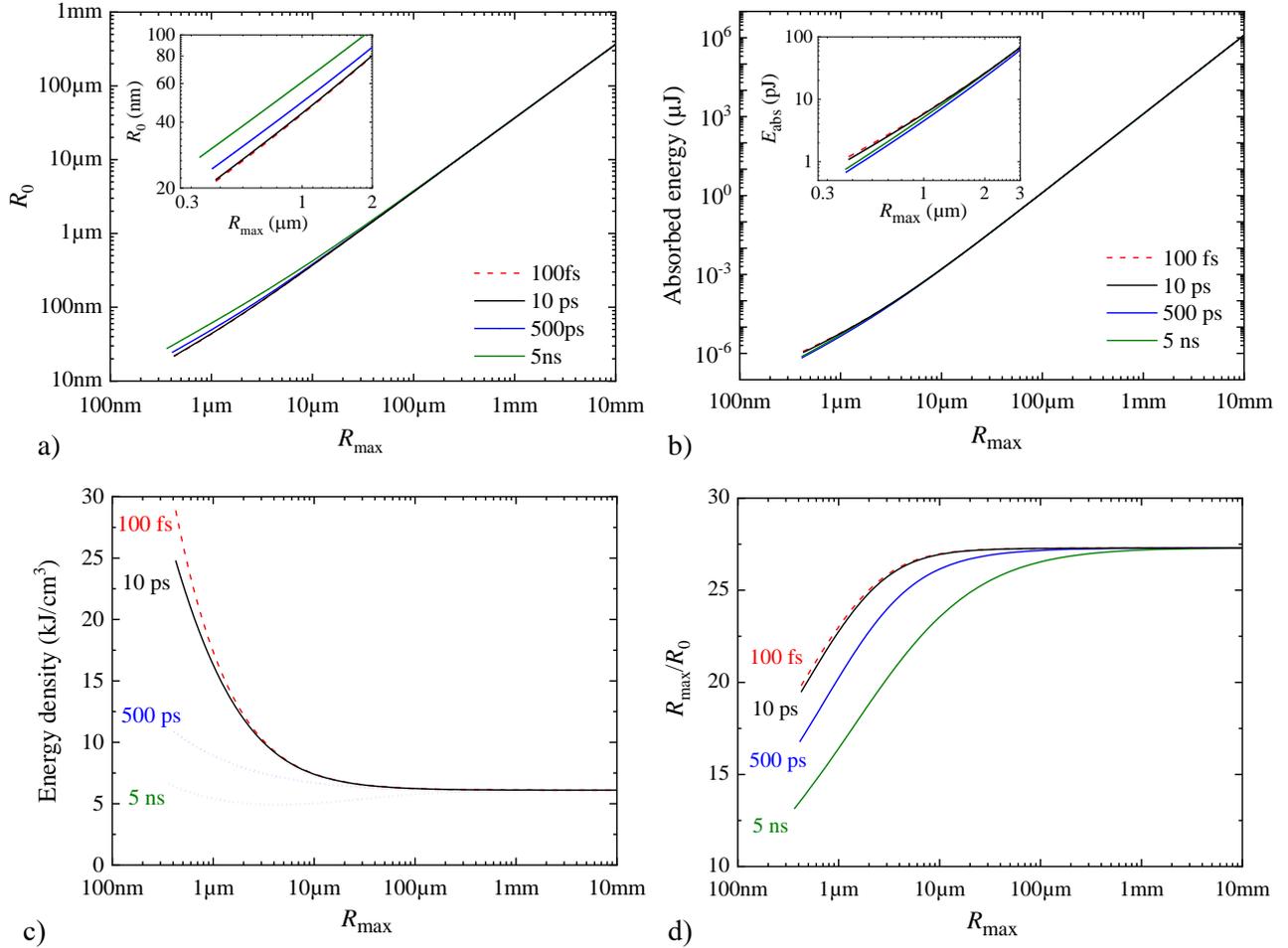

**FIG. 8** Bubble size dependence of (a) the initial bubble size $R_0$, (b) the absorbed energy $E_{abs}$ and (c) the energy density $\varepsilon = E_{abs}/(4/3\pi R_0^3)$ needed to produce $R_{max}$ at constant $R_{nbd}/R_0$ and different laser pulse durations. Curves for $\tau_L = 100$ fs are dashed because of the uncertainty of the thermalization time at small $\varepsilon_{abs}$. Curves for $\tau_L = 500$ ps and $\tau_L = 5$ ns in (c) are drawn as dotted lines because the $\varepsilon$ values refer to $R_0$ without considering the actual bubble enlargement beyond $R_0$ during the pulse. The bubble size increase relative to the seed bubble, $R_{max}/R_0$, is shown in (d). Simulations were performed for $R_{nbd}/R_0 = 10.4$.

Altogether, it becomes obvious that the ratio $R_{nbd}/R_0$ is a unique measure of $\varepsilon$ only when the inertial confinement condition is fulfilled and, additionally, bubbles that are so large that surface tension and viscosity become negligible.

Let us now proceed from an integral view of energy deposition during micro- and nanobubble formation to an analysis of the energy partitioning dynamics. Figure 9 compares the partitioning of deposited laser energy into internal energy, work done on the gas, work done against viscosity, and potential bubble energy for fs and ns breakdown at $R_{max} \approx 5$ µm. In all graphs, the 100% level refers to the deposited energy not required for vaporization, i.e. to ($E_{abs}$ - $E_v$) at the end of the laser pulse. With inertial confinement [100-fs pulse, Figs. 9(a) and (b)], all laser energy first goes into internal energy and is later partitioned into the different hydrodynamic channels, as described in Refs. [72,73,78]. By contrast, with longer pulses and lack of inertial confinement, energy deposition and partitioning go hand in hand. At $\tau_L = 5$ ns [Figs. 9(c) and (d)], only 58.5% of the deposited energy not needed for vaporization transiently takes the form of internal bubble energy. This conversion peaks during the second half of the laser pulse. At the end of the laser pulse (at $t = 2\tau_L$), 48% of ($E_{abs}$ - $E_v$) is already consumed by work done against the surrounding liquid.

Concurrency of energy deposition and conversion into mechanical energy lowers the strength of the mechanical effects. At $R = R_{max}$, the kinetic energy is zero and we can evaluate the conversion into acoustic energy using Eq. (28). For $R_{max} \approx 5$ µm, the fraction of total mechanical work during bubble expansion going into acoustic energy, $E_{acoust}/W_{gas,Rmax1}$, amounts to 43.84 % for the 100-fs pulse but to only 15.85 % for the 5-ns pulse.

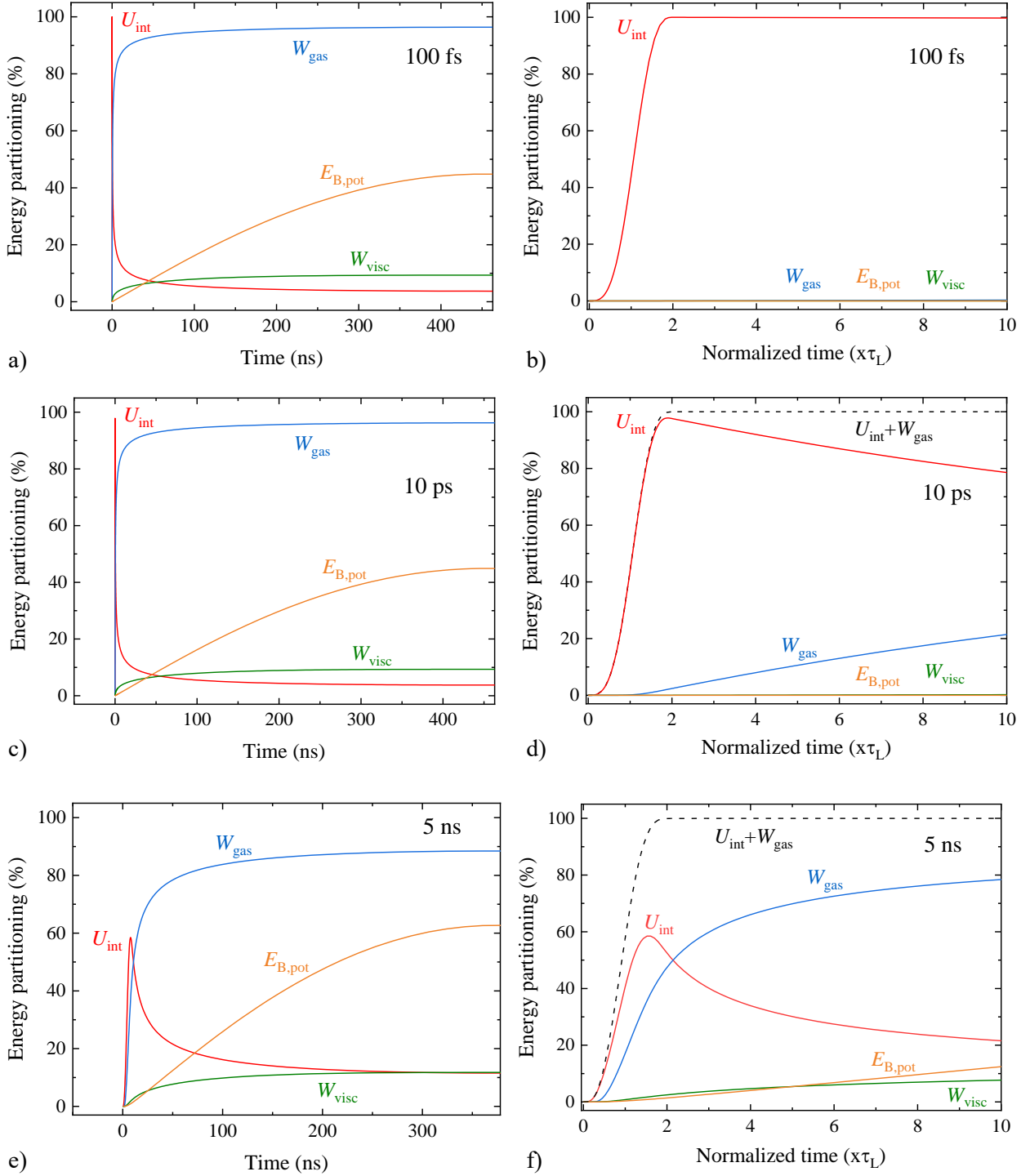

**FIG. 9** Time evolution of the internal bubble energy $U_{int}$ capable of doing work, the total amount of work $W_{gas}$ done on the liquid by the expanding gas, the fraction $W_{visc}$ done for overcoming viscosity, and the potential bubble energy $E_{B,pot}$ after laser-induced breakdown for 100-fs pulses (a), (b), 10-ps pulses (c), (d), and 5-ns pulses (e), (f). Simulations are performed for the same parameters as in Figs. 5 and 6 ($R_0 = 210$ nm and $R_{nbd} = 10.4 \, R_0$), which lead to a maximum bubble size of $R_{max} \approx 5 \, \mu m$ [exact values are indicated in Fig. 3(a)]. The entire time period up to maximum bubble expansion is presented in (a), (c) and (e), while (b), (d) and (f) show the initial expansion phase on a time scale normalized by the laser pulse duration. The 100% level always refers to $[U_{int}(t) + W_{gas}(t)]$, which is equivalent to $(E_{abs} - E_v)$ after the end of the laser pulse, as seen in (d) and (f).

Figure 10 presents the partitioning of absorbed laser energy into $E_v$, $E_{SW}$, $E_{B,pot}$ and $W_{visc}$ evaluated at $R_{max}$ for $R_{max} \rightarrow 0$ at different pulse durations. To understand the complex $R_{max}$ dependence, we first look at the self-similarity regime for $R_{max} > 100$ µm, where bubble formation is inertially confined and surface tension as well as viscosity are negligibly small. At $R_{max} = 1$ mm, 26.8% of the absorbed energy is needed for vaporization of the liquid in the plasma volume, 22.2% goes into acoustic transient emission, 50.9% into bubble formation, and 0.04% is needed to overcome viscous damping. Here, a larger fraction goes into bubble energy than into acoustic energy because with $R_{nbd}/R_0 = 10.4$ the plasma energy density and temperature are relatively small ($T = 1550$ K [46]). For larger $R_{nbd}/R_0$, the partitioning would shift towards acoustic energy. Linz et al. have shown that $E_{acoust}/E_{B,pot}$ strongly increases with plasma energy density [64].

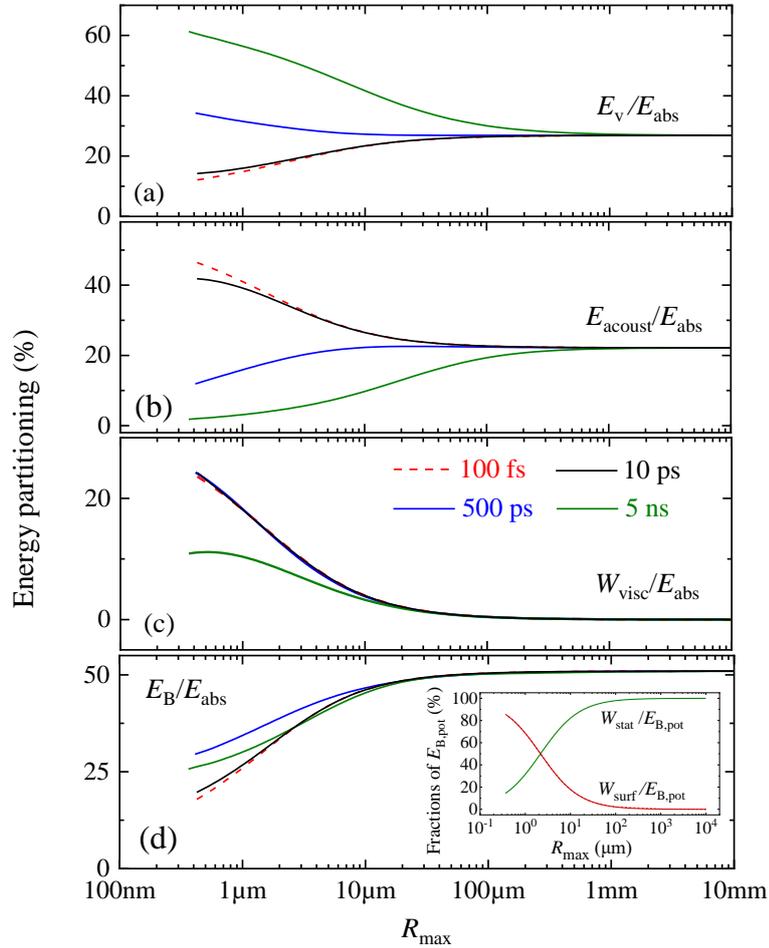

**FIG. 10** Bubble size dependence of the fractions of absorbed energy involved in producing a laser-generated bubble for different pulse durations. (a) Vaporization energy; b) energy of the shock wave or acoustic transient; (c) energy needed to overcome viscous damping during bubble expansion; (d) total energy of the expanded bubble according to Eq. (33). The insert in (d) shows the fractions of the potential energy $E_{B,pot}$ of the maximally expanded bubble that are related to the hydrostatic pressure ($W_{stat}/E_{B,pot}$) and to surface tension ($W_\sigma/E_{B,pot}$). A value $R_{nbd}/R_0 = 10.4$ was used in all simulations.

With decreasing bubble size, the energy partitioning changes dramatically and diverges for different laser pulse durations. Let us first look at $\tau_L = 100$ fs, where energy deposition remains inertially confined and only the influence of surface tension and viscosity changes. The increase of viscous damping is most pronounced, when the bubble wall velocity is largest, which is the case for the shortest pulse duration. Therefore, the fraction $W_{\mathrm{visc}}/E_{\mathrm{abs}}$ increases very strongly [Fig. 10(c)]. We have seen in Fig. 8(c) that overcoming viscous damping and surface tension requires an ever-larger energy density for $R_{\mathrm{max}} \rightarrow 0$. The growth of $\varepsilon$ is coupled with a decrease of the fraction $E_v/E_{\mathrm{abs}}$ going into vaporization and with an increase of the fraction $E_{\mathrm{acoust}}/E_{\mathrm{abs}}$ going into acoustic energy [Figs. 10(a) and (b)], consistent with the pressure rise for $R_{\mathrm{max}} \rightarrow 0$ at $\tau_L = 100$ fs in Fig. 7(a). By contrast, the fraction $E_B/E_{\mathrm{abs}}$ remaining as energy of the expanded bubble decreases [Fig. 10(d)] because ever more energy is consumed for overcoming viscous damping. The bubble energy $E_B$ consists mainly of potential energy $E_{\mathrm{B,pot}}$, together with a small amount of internal energy [Eq. (33)]. This applies regardless of bubble size, while the composition of $E_{\mathrm{B,pot}}$ changes strongly for $R_{\mathrm{max}} \rightarrow 0$, since an ever-larger fraction becomes related to surface tension [insert in Fig. 10(d)].

With longer pulse durations, not only the influence of viscosity and surface tension increases for $R_{\mathrm{max}} \rightarrow 0$ but also inertial confinement gets lost. The resulting features are most pronounced for $\tau_L = 5$ ns. Since the bubble expands during the laser pulse, the breakdown pressure is reduced and little energy is converted into acoustic energy [Fig. 10(b)]. Because of the smaller bubble wall velocity, less energy is lost by viscous damping than at 100 fs but this fraction is still very important [Fig. 10(c)]. Since viscous damping is reduced, the fraction going into bubble energy increases [Fig. 10(d)]. Altogether, the mechanical effects are mitigated by the loss of inertial confinement, and the fraction going into vaporization increases. It rises up to 60% at $R_{\mathrm{max}} = 0.5$ µm [Fig. 10(a)].

In order to assess changes of the character of the mechanical laser effects in dependence of inertial confinement, it is interesting to compare the conversion into acoustic and bubble energy. For small bubbles, an increase of laser pulse duration goes along with a lower conversion into acoustic energy, while the conversion into bubble energy is only weakly affected. We can read from Figs. 10(b) and (d) that at $R_{\mathrm{max}} \approx 5$ µm, the ratio $E_{\mathrm{acoust}}/E_{\mathrm{abs}}$ amounts to 29.35 % for $\tau_L = 100$-fs and to merely 6.65 % for $\tau_L = 5$-ns, while $E_B/E_{\mathrm{abs}}$ is 42.1% for the 100-fs pulse and 40.9% for the 5-ns pulse. Figure 11 presents the $E_{\mathrm{acoust}}/E_{\mathrm{B,pot}}$ ratio as a function of $\tau_L$ and $R_{\mathrm{max}}$. The ratio depends strongly on bubble size and pulse duration. For $R_{max} = 0.5$ μm, $E_{acoust} / E_{B,pot}$ is 30 times larger for fs breakdown than for ns breakdown, while for large bubbles and inertially confined energy deposition at all pulse durations it converges to a value of 0.665.

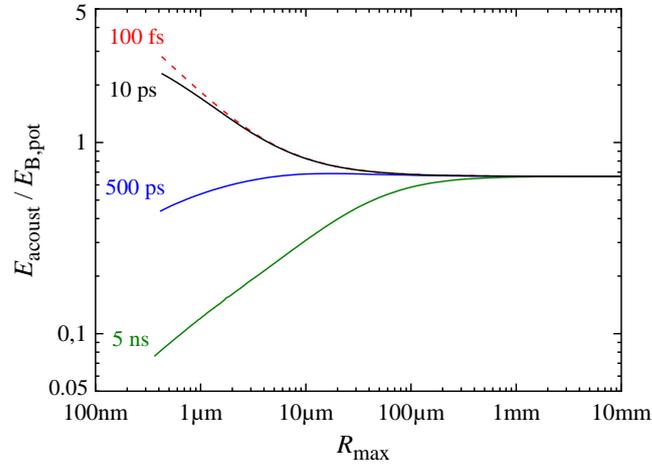

**FIG. 11** Ratio of acoustic energy $E_{acoust}$ to the potential energy $E_{B,pot}$ of the expanded bubble plotted as a function of bubble size for different pulse durations and $R_{nbd}/R_0 = 10.4$.

In toto, we see that mechanical effects during nano- and microbubble formation with ultrashort laser pulses are closely linked to acoustic transient emission, while they are much gentler with ns pulses, where most of the mechanical energy goes into the surface energy of the expanded bubble. In the next section, we analyze under which circumstances a shock front forms, and how the shock pressure depends on laser pulse duration.

## VI. ACOUSTIC TRANSIENT EMISSION AND SHOCK FORMATION

Figure 12 presents the acoustic transient emission during the generation of a bubble with $R_{max} \approx 5$ µm for different laser pulse durations. The $\tau_L$ dependence of the respective peak pressures and particle velocities in the liquid and at the shock front is summarized in Fig. 13.

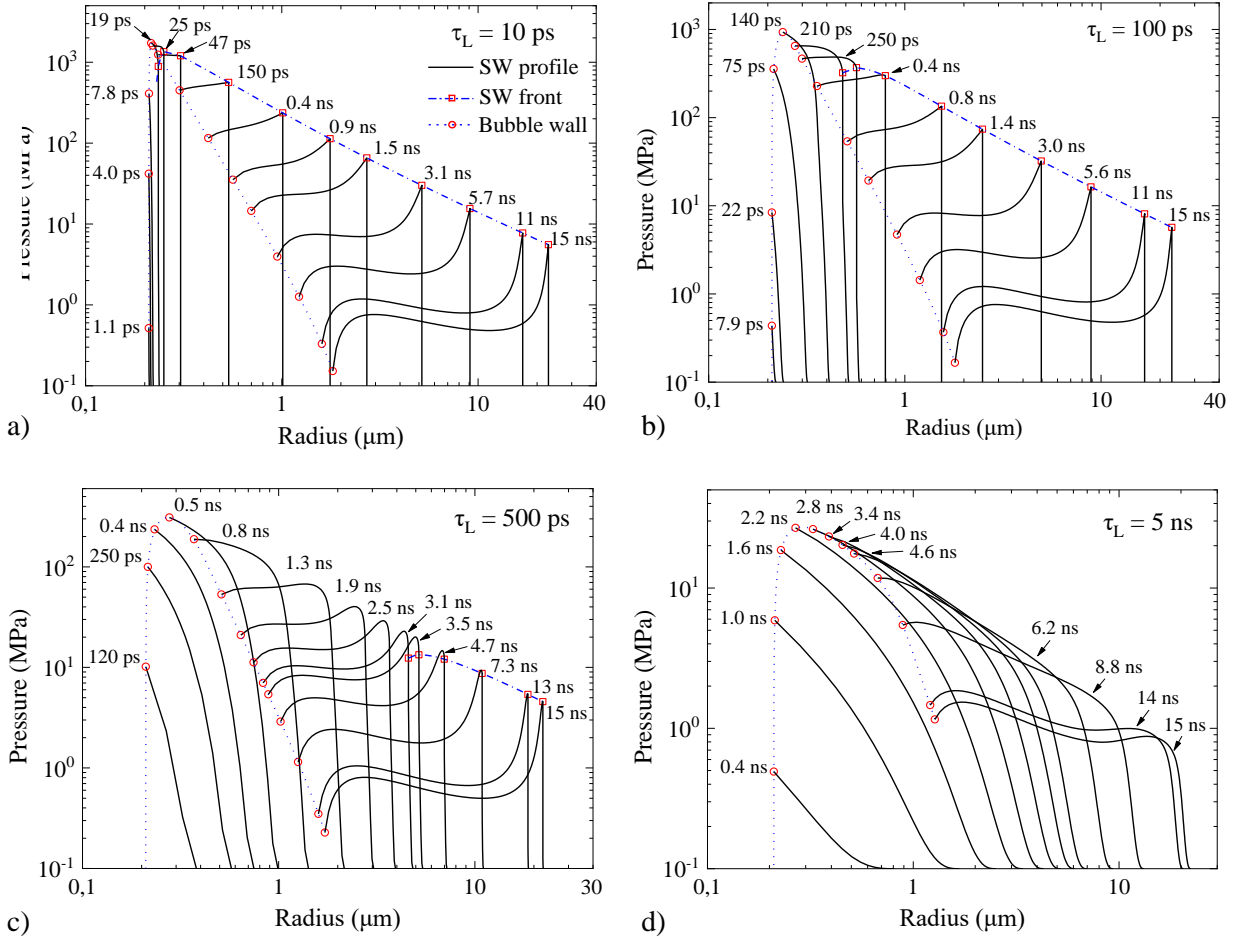

**FIG. 12** Evolution of the pressure distribution $p(r)$ during acoustic transient emission after optical breakdown for (a) $\tau_L = 10$ ps, (b) 100 ps, (c) 500 ps and (d) 5 ns. All simulations were performed for $R_0 = 210$ nm and $R_{nbd} = 10.4 \times R_0$ and cover the first 15 ns after the start of the laser pulse. The maximum radius values in (a) to (d) are 5.57 µm, 5.50 µm, 5.30 µm and 4.51 µm, respectively, like in Figs. 3, 5, 6, and 9. The location of the bubble wall and the respective pressure values are indicated by circles, and the location of the pressure peak at the shock front is indicated by squares. For the 100-fs pulse, the shock front forms immediately after optical breakdown and the maximum pressure in the liquid, $p_{max}$, approximately equals the maximum pressure jump at the shock front, $\Delta p_{sw,max}$. For $\tau_L = 100$ ps, the shock front forms after 250 ps and for $\tau_L = 500$ ps, it forms only after 3.5 ns. With increasing pulse duration $\Delta p_{sw,max}$ decreases, and with the 5-ns pulse, no shock front forms.

Peak pressure and velocity in the liquid decrease with increasing pulse duration, as already found for the pressure and velocity at the bubble wall (Fig. 6). After ultrashort laser pulses, a shock wave evolves but for $\tau_L = 5$ ns, no shock front develops at all. The pressure jump at shock front decreases from $\Delta p_{sw,max} = 1924$ MPa at $\tau_L = 100$ fs to 13.4 MPa at $\tau_L = 500$ ps, and a similar decrease is observed also for the particle velocity behind the shock front (from 671 m/s to 8.8 m/s). Altogether, the breakdown dynamics is strongly disruptive for fs pulses but much gentler for ns pulses.

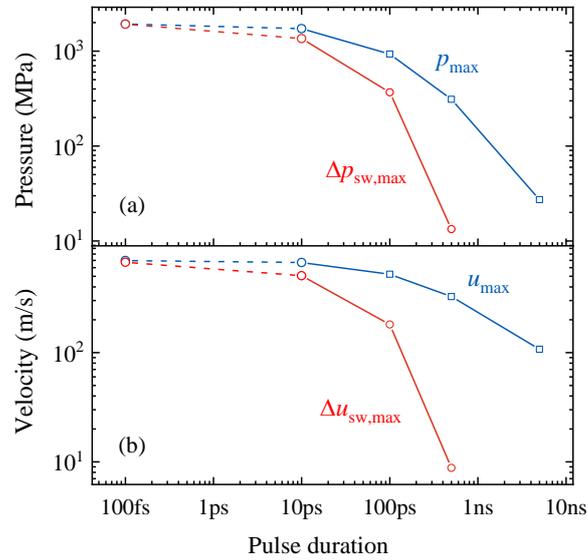

**FIG. 13** (a) Pulse duration dependence of the peak pressure in the liquid, $p_{max}$, and the maximum pressure jump at shock front, $\Delta p_{sw,max}$. (b) Pulse duration dependence of the peak velocity within the liquid, $u_{max}$, and the maximum particle velocity behind the shock front, $\Delta u_{sw,max}$. The respective $R_{max}$ values are the same as in Fig. 12. Between $\tau_L = 100$ fs and $\tau_L = 10$ ps, the lines are dashed lines because of the uncertainty of the thermalization time at small $\varepsilon_{abs}$.

# VII. CONSEQUENCES FOR LASER SURGERY AND MICROFLUIDICS

Laser surgery and laser-induction of microfluidic streaming often involve small bubbles in the micrometer or even nanometer range. We have seen above that inertial confinement is progressively lost with increasing laser pulse duration and decreasing bubble size, when significant bubble expansion starts already during the laser pulse. Only for fs pulses, energy deposition is fully confined and hydrodynamic events set in after the energy deposition is completed. However, even here, the dynamics changes for $R_{max} \to 0$ because the influence of surface tension and viscosity increases with decreasing bubble size.

The high degree of inertial confinement for ultrashort laser pulses results in particularly large breakdown pressures and initial bubble wall velocities during micro- and nanobubble formation. Furthermore, the ratio $E_{acoust}/E_{B,pot}$ between acoustic and bubble energy increases with decreasing pulse duration and bubble size, and a shock front evolves only during micro- and nanobubble formation with ultrashort pulse durations but not during ns breakdown. Therefore, microeffects produced by fs laser pulses are more disruptive than those from ns breakdown.

Our simulations for spherical bubble dynamics correspond to optical breakdown at large numerical aperture ($NA$) with compact plasmas. Here disruptiveness increases with decreasing laser pulse duration, while the opposite trend has been observed at moderate $NA$ [56]. For breakdown in water produced at $0.15 \leq NA \leq 0.25$, plasma energy density and conversion into mechanical energy were smaller for fs pulses than for ns pulses, especially at pulse energies well above threshold. The opposite trends at large and small numerical apertures can be explained by different movements of the optical breakdown wave during the laser pulse. At large $NA$s, the physical plasma length is short and the breakdown wave moves always upstream from the beam waist towards the incoming laser beam. This produces a large plasma energy density, regardless of pulse durations. At small $NA$s and large pulse energies, the plasma region in fs breakdown is longer than the length of the laser pulse. In this case, the breakdown wave propagates with the laser pulse towards the beam waist, which goes along with continuous energy depletion and a reduction of plasma energy density [64,79,80]. Filamentation prolongs the plasma region further and diminishes its energy density [81]. By contrast, for ps and ns breakdown, the breakdown wave moves upstream also at small $NA$s, and plasma energy density remains high. We conclude that the focusing angle is an important control parameter not only for the plasma shape and length but also for the pulse duration dependence of the disruptiveness of breakdown events.

Figures 10 and 11 indicate that the increase of disruptiveness with decreasing $\tau_L$ becomes ever-more pronounced with decreasing bubble size. It should be noted, however, that our simulations for $R_{max} \to 0$ can only elucidate general trends because we assumed a constant $R_{nbd}/R_0$ value in all simulations. Therefore, the predicted trends are accurate only in the pulse energy range where the plasma energy density is approximately constant. In reality, the generation of ever smaller bubbles goes along with decreasing laser pulse energies and a decrease of plasma energy density. Nevertheless, our results on the bubble size dependence of bubble dynamics and acoustic transient emission are realistic for most practical cases, where super-threshold pulse energies are used. Close to $E_{th}$, the bubble size increases very fast with pulse energy, and at $E_L = 1.2\ E_{th}$, the radius of bubbles produced by IR fs laser pulses is already more than ten times larger than at threshold [33,64]. Thus, on a pulse energy scale, the transition zone from bubble threshold to constant plasma energy density is very small, and our assumptions of approximately constant $\varepsilon$ are probably valid for $E_L \geq 1.2\ E_{th}$. In this regime, thermalization times in the ps range are expected as discussed in section II.C, and the simulation results for $\tau_L = 10$ ps seem adequate. By contrast, for nano-and microbubble generation in the range $E_L < 1.2\ E_{th}$, simulation results for $\tau_L = 100$ fs will be closer to reality.

For an accurate coverage of the nanobubble regime close to threshold, further studies based on experimental data on the pulse energy dependence of $R_{max}$, $R_0$ and plasma absorption are needed. Measurements of $R_0$ are challenging because fs and ps plasmas close to the bubble threshold luminesce very weakly. Moreover, $R_0$ may be below the optical resolution limit, where only light scattering methods provide sufficient spatial resolution [33]. Close to $E_{th}$, the plasma absorption is very small and scattering may cause far stronger changes of plasma transmission than absorption. This makes it impossible to determine the absorbed laser energy through simple transmission measurements. In principle, $E_{abs}$ can be determined through optical breakdown modeling but appropriate models suitable for large focusing angles and different pulse durations still need to be developed.

Our results for water at $R_{max} \approx 5$ μm suggest that major changes of inertial confinement and breakdown dynamics occur, when the pulse duration is reduced from 5 ns to 100 ps, whereas further shortening of $\tau_L$ brings only a gradual increase of shock pressure and bubble wall velocity (Figs. 6, 12, 13). However, in experiments on corneal laser dissection, we observed a decrease of the energy threshold by one order of magnitude, when the pulse duration was reduced from > 200 ps to ≤ 1 ps [7,82].

Apparently, the increase of disruptiveness of breakdown effects in viscoelastic media is shifted towards shorter pulse durations compared to water. It seems reasonable that the disruptiveness of optical breakdown leading to a given bubble size is altered by the viscoelastic material response. Already in water, $E_{acoust}/E_{abs}$ and $E_{acoust} / E_{B,pot}$ grow with decreasing bubble size for ultrashort laser pulses due to the increasing influence of surface tension and viscosity (Figs. 10 and 11). The deviatoric stress arising from tissue elasticity has a similar but stronger effect on bubble oscillations as surface tension in water, and the viscosity of tissue is also larger than that of water. Therefore, the increase of disruptiveness in laser-induced bubble generation with decreasing pulse duration and bubble size should be more pronounced in tissue than in water. For testing this hypothesis, the radial deviatoric stress corresponding to the viscoelastic properties of the medium, as well as the rupture strength of the medium must be included into the bubble model [83-85].

The modeling tools presented in this paper are applicable not only to free-focused laser pulses but also to bubble formation around micro- and nanoparticles. Femtosecond irradiation of metallic nanoparticles at off-resonance wavelengths produces plasma in a thin liquid layer in the field-enhanced region just outside the particle [86-88]. This leads to a sudden pressure rise, and the bubble wall will feature a jump-start like for the case of plasma formation in free liquid. The Gilmore model can be used to simulate laser-induced bubble generation by treating the NP as a van der Waals hardcore surrounded by a thin shell driving the bubble expansion. Modeling is more complex for particles featuring strong linear absorption (e.g. melanin granules) or for plasmonic nanoparticles excited at wavelength around the plasmon resonance. Here, the particle itself is the primary heat source and bubble formation is mediated by heat diffusion into the surrounding water [47,89-92]. Thus, the convolution of laser pulse duration with the particle's thermal relaxation time plays a similar role as the laser pulse duration alone for plasma-mediated energy deposition. Consequently, inertial confinement is more easily lost and bubble formation around nanoparticles is likely gentler for visible laser wavelengths near the plasmon resonance than with ultrashort laser pulses at off-resonant IR wavelengths. Thus, while the disruptiveness of mechanical effects by free-focused laser pulses can be adjusted through the choice of the laser pulse duration, the disruptiveness of nanoparticle-mediated mechanical effects can be adjusted by tuning the laser wavelength. This topic deserves further numerical and experimental investigations.

Our findings on the influence of inertial confinement on laser-induced bubble generation and shock wave emission are relevant also for numerical simulations using finite volume approaches [16,17,20]. Here, the interlacing between pressure distribution and particle velocity in finite amplitude waves, which we incorporated into a spherical bubble model, is automatically included. However, most finite volume studies still assume instantaneous, isochoric deposition of the energy creating the cavitation bubbles, which is not appropriate for laser bubbles outside the inertial confinement regime. Thus, the time course of energy deposition should be considered also in future finite volume studies, especially if they are applied to micro- and nanocavitation. Such models then bear the potential to consider an aspherical shape of the laser plasma, spatial anisotropies within the plasma (e.g. a hot spot with large energy density), and asymmetric boundary conditions.

# VIII. CONCLUSIONS AND OUTLOOK

Laser-generated micro- and nanobubbles expand significantly during a ns laser pulse, and energy deposition is isochoric and fully inertially confined only for ultrashort laser pulses. This difference results in much larger breakdown pressures and initial bubble wall velocities with ultrashort laser pulses than for longer pulse durations. Furthermore, a shock front evolves only with ultrashort pulses but not during ns breakdown. Therefore, fs breakdown is more disruptive than ns breakdown producing bubbles of equal size. This has important implications for laser surgery of cells and tissues because the disruptiveness can be adjusted through the choice of the laser pulse duration. To capture these features, the finite duration of the laser pulses and the evolution of particle velocity at the bubble wall resulting from nonlinear sound propagation at bubble pressure $P$ must be considered. The extended Gilmore model presented in Ref. [46] with its generalization in the present paper achieves this task for any degree of inertial confinement. Similar features should be integrated also in other advanced bubble models designed to simulate laser-induced bubble formation and shock wave emission [93,94].

In this paper, we have investigated the influence of inertial confinement on laser-induced bubble dynamics in water. For simulating bubble dynamics in cells and tissues, the viscoeleastic properties of the optical breakdown medium and its rupture strength have to be included [83,84]. Such features can be integrated into the extended Gilmore model if the mechanical properties under high strain rates are known, which differ largely from the properties under physiological conditions [48]. The mechanical properties can be obtained iteratively by comparing model predictions to experimental results on laser-induced bubble dynamics and acoustic transient emission [7,26,41,44,83,95].

In our simulations, the ambient pressure was kept constant at 0.1 MPa, which is representative for most applications of laser-induced bubble formation. However, laser-induced cavitation at elevated pressure is relevant for laser ablation in pressurized liquids [96,97] and for deep-sea laser-induced breakdown spectroscopy (LIBS) [52,53]. Both applications often involve the use of energetic nanosecond laser pulses, for which bubble expansion starts during the laser pulse. An elevated ambient pressure increases the confinement and slows the expansion. That leads to a higher breakdown pressure and delayed adiabatic cooling, which can influence ablation dynamics and LIBS signal. The changes of shock wave emission and bubble dynamics in laser ablation and LIBS at elevated pressure can be well portrayed by the model presented in this paper.

We have mostly assumed a moderate plasma energy density, represented by $R_{nbd}/R_0 = 10.4$, which corresponds to a plasma temperature around 1500 K [46]. Here, condensation during bubble expansion progresses quickly and the net condensation during the first oscillation cycles can be considered by changing $R_n$ at $R_{max}$ such the period of the second oscillation matches experimental data [24,46]. However, with decreasing energy density, an ever larger part of the absorbed energy is needed for vaporization, and the bubble expansion slows down, especially for long pulse durations. Close to the bubble threshold, condensation of the vaporized liquid in the plasma may not be complete before $R_{max1}$ is reached. Under these circumstances, bubble dynamics becomes thermally driven [98], and models must continuously track evaporation and condensation during the bubble oscillations [61,66,99-101]. Combination of such tracking with the consideration of accelerated bubble wall movement during acoustic transient emission remains a task for future work.

A major challenge is the comparison of numerical predictions with experimental data because the changes of inertial confinement are most pronounced for micro- and nano-cavitation. With decreasing bubble size, the requirements on spatiotemporal resolution increase and the space of parameters influencing bubble dynamics becomes larger (Fig. 1). Light scattering techniques enable single-shot acquisition of $R(t)$ data and bubble oscillation times with very high precision [33,34,37,46,102], while speckle-free imaging with sub-micrometer spatial and sub-nanometer temporal resolution can capture the acoustic transient emission during microbubble generation and collapse. For bubble dynamics in water featuring high reproducibility, this can be achieved by stroboscopic photography [10], and for breakdown in cells and tissue, multi-exposure single-frame photography [103-106] may be a solution, if the pulse repetition rate can be increased to ≈1 GHz.

A deeper understanding of the influence of inertial confinement on laser-induced bubble generation and shock wave emission enables a better control of optical breakdown events in laser surgery, microfluidics, and material processing in liquid environments. The present study has pointed out its importance, provided a modeling tool and first insights but future theoretical and experimental studies are warranted.

# ACKNOWLEDGMENT

This work was supported by U.S. Air Force Office of Scientific Research, grant FA9550-22-1-0289.

# AUTHOR DECLARATIONS

## Conflict of Interest

The authors have no conflicts to disclose.

## Author contributions

**Xiao-Xuan Liang:** Conceptualization (equal); Formal analysis (lead); Software (lead); Data interpretation (equal), Writing – review & editing (equal).

**Alfred Vogel:** Conceptualization (equal), Fund raising (lead), Data interpretation (equal), Writing – original draft (lead); Writing – review & editing (equal).

# DATA AVAILABILITY

Data sharing is not applicable to this article as all results can be reproduced with the used open-source LIBDAR software, as referenced in section II.C.

# APPENDIX: GENERALIZED TIME DERIVATIVE OF BUBBLE WALL PRESSURE

The time-derivative of bubble wall pressure for the general case of bubble expansion without inertial confinement [Eq. (22)] is obtained starting with Eq. (19). For the sake of simplicity, we replace here the time-dependent quantities $P(t)$, $R_n(t)$ and $R(t)$ by the symbols $P$, $R_n$ and $R$. Rearrangement of the terms on the right-hand side of Eq. (19) then leads to

$$P = p_\infty R_n^4 R^{-4} + 2\sigma R_n^3 R^{-4}. \tag{A1}$$

Temporal derivation of both sides leads to

$$\dot{P} = p_\infty \frac{d\left(R_n^4\right)}{dt} R^{-4} + p_\infty R_n^4 \frac{d\left(R^{-4}\right)}{dt} + 2\sigma \frac{d\left(R_n^3\right)}{dt} R^{-4} + 2\sigma R_n^3 \frac{d\left(R^{-4}\right)}{dt}, \tag{A2}$$

and calculation of the derivatives of the functions in the brackets yields

$$\dot{P} = 4 p_\infty R_n^3 \dot{R}_n R^{-4} - 4 p_\infty R_n^4 R^{-5} \dot{R} + 6 \sigma R_n^2 \dot{R}_n R^{-4} - 8 \sigma R_n^3 R^{-5} \dot{R} \,. \tag{A3}$$

Through some rearrangements, Eq. (A3) can be condensed to

$$\dot{P} = 4 \left( p_\infty + \frac{2\sigma}{R_n} \right) \frac{R_n^4}{R^4} \left( \frac{\dot{R}_n}{R_n} - \frac{\dot{R}}{R} \right) - \frac{2\sigma \dot{R}_n R_n^2}{R^4} \,, \tag{A4}$$

which corresponds to Eq. (22) in the main text in compact notation.

Using the same approach, one can derive from Eq. (8) the temporal derivative of bubble wall pressure for arbitrary polytropic exponent of the gas in the bubble:

$$\dot{P} = 3\kappa \left( p_\infty + \frac{2\sigma}{R_n} \right) \frac{R_n^{3\kappa}}{R^{3\kappa}} \left( \frac{\dot{R}_n}{R_n} - \frac{\dot{R}}{R} \right) - \frac{2\sigma \dot{R}_n}{R_n^2} \frac{R_n^{3\kappa}}{R^{3\kappa}} \,. \tag{A5}$$